\documentclass[aps,prb,twocolumn,floatfix,amsmath,amssymb]{revtex4}
\usepackage{graphicx}
\usepackage{dcolumn}
\usepackage{bm}
\usepackage{times}
\renewcommand{\vec}{\mathbf}

\newcommand{\me}{\mathrm{e}}
\newcommand{\mi}{\mathrm{i}}

\begin{document}

\title{Quantum oscillations in electron doped high temperature superconductors}

\author{Jonghyoun Eun}
\author{Xun Jia}
\author{Sudip Chakravarty}
\affiliation{Department of Physics and Astronomy, University of
California Los Angeles, Los Angeles, California 90095-1547, USA}
\date{\today}

\begin{abstract}
 {Quantum oscillations in hole doped high temperature superconductors are difficult to understand within the prevailing views. An emerging idea is that of a putative normal ground state, which appears to be a Fermi liquid with a reconstructed Fermi surface. The oscillations are due to formation of Landau levels. Recently the same oscillations were found  in the electron doped cuprate, $\mathrm{Nd_{2-x}Ce_{x}CuO_{4}}$,  in the optimal to overdoped regime. Although these electron doped non-stoichiometric materials are  naturally more disordered, they strikingly complement the  hole doped cuprates. Here we  provide  an explanation of these observations from the perspective of density 
waves using a powerful transfer matrix method to compute the conductance as a function of the magnetic field.}  
\end{abstract}

\maketitle 
\section{Introduction}
 Periodically new experiments tend to disturb the status quo of the prevailing  views in the area of high temperature cuprate superconductors.  Recent quantum oscillation (QO) experiments~\cite{Doiron-Leyraud:2007,Bangura:2008,LeBoeuf:2007,Jaudet:2008,Yelland:2008,Sebastian:2008,Audouard:2009,Singleton:2009} fall into this category.~\cite{Chakravarty:2008} The first set of experiments were carried out in underdoped high quality crystals of well-ordered $\mathrm{YBa_{2}Cu_{3}O_{6+\delta}}$ (YBCO), stoichiometric  $\mathrm{YBa_{2}Cu_{4}O_{8}}$ (Y124) and the overdoped single layer $\mathrm{Tl_{2}Ba_{2}CuO_{6+\delta}}$~\cite{vignolle:2008}. 
 
More recently oscillations are also observed in electron doped $\mathrm{Nd_{2-x}Ce_{x}CuO_{4}}$ (NCCO).~\cite{Helm:2009}  The measurements in NCCO  for 15\%, 16\%, and 17\% doping~\cite{Helm:2009} are spectacular.   The salient features are: (1) The experiments are performed in the range $30-64 T$, far above
the upper critical field, which is about $10 T$ or less; (2) the material involves single CuO plane, and therefore
complications involving chains, bilayers, Ortho-II potential,~\cite{Podolsky:2008} etc.  are absent; (3) stripes~\cite{Millis:2007} may not be germane
in this case.~\cite{Armitage:2009} It is true, however, that neither  spin density wave (SDW) nor $d$-density wave (DDW)~\cite{Chakravarty:2001} are yet directly observed  in NCCO in the relevant doping range, but QOs  seem to require their existence, at least the {\em field induced} variety (see, however Ref.~\onlinecite{Rourke:2009}); (4) these experiments are a tour de force because the sample is
non-stoichiometric with naturally greater intrinsic disorder. The  effect is therefore no longer confined to a limited class
of high quality single crystals; (5) The authors have also succeeded in seeing the transition from
low to high frequency oscillations~\cite{Kusko:2002} in NCCO as a function of doping. 

Here we focus on  NCCO.  We shall see that disorder plays an important role. Without it it is impossible to understand why the slow oscillations damp out below $30 T$ for 15\% and 16\% doping, and below $60 T$ for 17\% doping, even though the field range is very high. For 17\% doping, where a large hole pocket is observed corresponding to very fast oscillations (inconsistent with any kind of density wave order), the necessity of such   high fields can have only one explanation, namely to   achieve a sufficiently large $\omega_{c}\tau$, where $\omega_{c}=eB/m^{*}c$, $\tau$ is the scattering lifetime of the putative normal phase, $m^{*}$ the effective mass, and $B$ the magnetic field.  Qualitatively, the Dingle factor, $D$,  that suppresses quantum oscillations is $D=e^{-p\pi/\omega_{c}\tau}$
where $p$ is the index  for the harmonic. Assuming a Fermi velocity, suitably averaged over an orbit to be  $v_{F}$, the  mean free path $l=v_{F}\tau$. Thus  $D$ can be rewritten as $D= e^{-p\pi\hbar c k_{F}/eBl}$. A crude measure for  $k_{F}$ is given by expressing the area of an extremal orbit, $A$, as $A=\pi k_{F}^{2}$. By setting $m^{*}v_{F}=\hbar k_{F}$ the explicit dependence on the  parameters $m^{*}$ and $v_{F}$ was eliminated.
Assuming that the mean free paths for the hole and the electron pockets are more or less the same, not an unreasonable assumption, the larger pockets, with larger $k_{F}$, will be strongly suppressed for the same value of the magnetic field because of the exponential sensitivity of $D$ to the pocket size. This argument is consistent with our  exact transfer matrix calculation  using  the Landauer formula for the conductance presented below.  

Here we show that the oscillation experiments in NCCO reflect a broken translational symmetry~\cite{Chakravarty:2008b} that reconstructs the Fermi surface in terms of electron and hole pockets.~\cite{Chakravarty:2008} The emphasis is not the transfer matrix method itself, but its use in explaining a major experiment in some detail. We study both SDW and singlet  DDW orders with the corresponding mean field Hamiltonians.  A more refined calculation, beyond the scope of the present paper, will be necessary to see the subtle distinction  between the two order parameters. 

In Sec. II we introduce our mean field Hamiltonians and in Sec. III we discuss the transfer matrix method for the computation of quantum oscillations of the conductance. Sec. IV contains the results of our numerical computations and Sec. V our conclusions.

\section{Mean field Hamiltonian}
We suggest that the experiments in NCCO can be understood from a suitable normal state because the applied magnetic fields between 30-65 T are so far above the upper critical field, which is less than 10 T, that  vortex physics and the superconducting gap  are not important. Our assumption is that a broken translational  symmetry state with an ordering vector ${\bf Q}=(\pi/a,\pi/a)$ ($a$ being the lattice spacing) can reconstruct the Fermi surface resulting in two hole pockets and one electron pocket within the reduced Brillouin zone, bounded by the constraints on the wave vectors $k_{x}\pm k_{y}=\pm \pi/ a$. One challenge here is to understand why the large electron pockets corresponding to 15 and 16\% doping  resulting from the band structure parameters for NCCO defined below are not observed, but the much smaller hole pockets are. Another challenge is to understand why the large Fermi surface at 17\% doping is not observed until the applied field reaches about 60 T. The reason we believe is the existence strong cation disorder in this material. It is therefore essential to incorporate disorder in our Hamiltonian. For the Hamiltonian itself, we consider a mean field approach, and for this purpose we consider two possible symmetries, one that corresponds to a singlet in the spin space (DDW) and one that is a triplet in the spin space (SDW). Note that these are particle-hole condensates for which orbital function does not constrain the spin wave function unlike a particle-particle condensate (superconductor) because there are no exchange requirements between a particle and a hole.

\begin{figure}
\begin{center}
\includegraphics[scale=0.5]{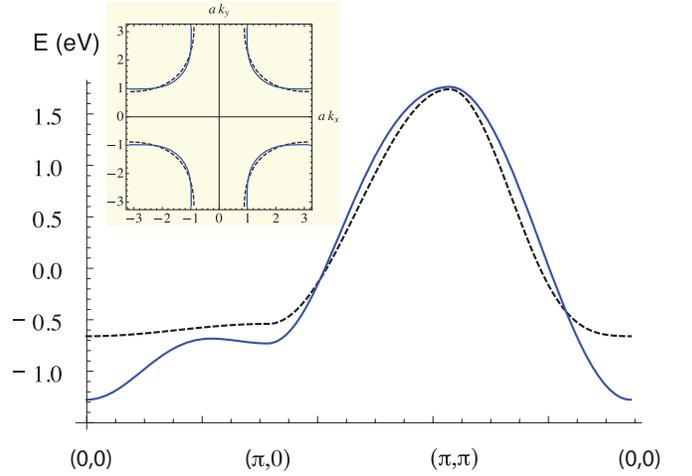}
\caption{(Color online) The solid  curve represents the $t-t'-t''$ band structure ($t=0.38 \mathrm{eV},\;  t'=0.32 t, \; t''=0.5 t'$), and the dashed curve corresponds to $t-t'$ band structure, (see Table~\ref{table1}). The quasiparticle energy is plotted in the Brillouin zone along the triangle $(0,0) \to (\pi,0) \to (\pi,\pi) \to (0,0)$. In the inset the chemical potential,  $\mu$, was adjusted to obtain approximately 15\% doping.}
\label{fig:band}
\end{center}
\end{figure}

We believe that it is reasonable  that as long as a system is deep inside a broken symmetry state, mean field theory and its associated elementary excitations should correctly capture the physics. The fluctuation effects will be important close to quantum phase transitions. However, there are no indications in the present experiments that fluctuations are important. 
The microscopic basis for singlet DDW Hamiltonian is discussed in some detail in Refs.~\onlinecite{Dimov:2008,Jia:2009} and in references therein. So, we do not see any particular need to duplicate this discussion here. 
The mean field Hamiltonian for the singlet DDW in real space, in terms of the site-based fermion annihilation and creation operators of spin $\sigma$, $c_{\vec{i},\sigma}$ and  $c_{\vec{i},\sigma}^{\dagger}$, is 
  \begin{equation}
                H_{DDW}=\sum_{\vec{i},\sigma}\epsilon_\vec{i}c_{\vec{i},\sigma}^\dag c_{\vec{i},\sigma}+\sum_{
                \vec{i},\vec{j},\sigma}t_{\vec{i},\vec{j}}~\mathrm{e}^{ia_{\vec{i},\vec{j}}}c_{\vec{i},\sigma}^\dag
                c_{\vec{j},\sigma}+h.c.,
            \end{equation}
where the nearest neighbor hopping matrix elements are
            \begin{eqnarray}
                t_{\vec{i},\vec{i}+\hat{\vec{x}}}&=&-t+\frac{i W_0}{4}(-1)^{(\vec{i_x}+\vec{i_y})}, \\
                t_{\vec{i},\vec{i}+\hat{\vec{y}}}&=&-t-\frac{i W_0}{4}(-1)^{(\vec{i_x}+\vec{i_y})},
            \end{eqnarray}
Here $W_{0}$ is the DDW gap. We also include the next nearest neighbor hopping  $t'$, whereas the third neighbor hopping $t''$ is ignored to simplify computational complexity without losing the essential aspects of the problem. The parameters $t$ and $t'$ are chosen  (see Table~\ref{table1}) to closely approximate the more conventional  band structure, as shown in Fig.~\ref{fig:band}. We have checked that the choice $t''=0$ provides reasonably consistent results for the frequencies in the absence of disorder. For example, for DDW, and $15\%$ doping, the hole pocket frequency is 185 T, and the corresponding  electron pocket frequency is 2394 T. 
\begin{table}[htdp]
\caption{The band parameters, the chemical potential, and the mean field parameters for DDW and SDW  used in our calculation. $F$ in Tesla corresponds to the calculated oscillation frequencies of the hole pocket, the so-called slow frequencies. The measured $F$ for 15\% doping is $290\pm 10$ T and for 16\% doping is $280\pm 15$ T. The calculated magnitude of $F$ does depend on the neglected $t''$.}
\begin{center}\begin{tabular}{|cccccccc|}\hline
Order & $t$ (eV) & $t'$ & $W_{0}$ & $V_{S}$ & $\mu$ & $V_{0}$ &$F$ (T) \\
\hline\hline
DDW 15\% & 0.3 & $0.45 t$ & $0.1 t$ & * & $-0.40 t $ & $0.8 t$ & 195 \\
DDW 16\% & 0.3 & $0.45 t$ & $0.1 t$ & * & $-0. 365 t$& $0.8 t$ & 165 \\
SDW 15\% & 0.3 & $0.45 t$ & * & $0.05 t $ & $-0.403 t$ & $0.8 t$ & 195 \\
SDW 16\% & 0.3 & $0.45 t$ & * & $0.05 t$ & $-0.366 t$ & $0.8 t$ & 173\\
\hline
\end{tabular} 
\end{center}
\label{table1}
\end{table}

Similarly, the SDW mean field Hamiltonian is
            \begin{equation}\begin{split}
             H_{SDW}&=\sum_{\vec{i},\sigma}\left[\epsilon_\vec{i}+\sigma V_{S} (-1)^{\vec{i_x}+\vec{i_y}}\right]
                c_{\vec{i},\sigma}^\dag c_{\vec{i},\sigma}\\
                &+\sum_{\vec{i},\vec{j},\sigma}t_{\vec{i},\vec{j}}~\mathrm{e}^{ia_{\vec{i},\vec{j}}}
                c_{\vec{i},\sigma}^\dag c_{\vec{j},\sigma}+h.c.
                \end{split}
            \end{equation}
and the spin $\sigma=\pm1$, while the magnitude of the SDW amplitude is $V_{S}$.   In both cases a
constant perpendicular magnetic field $B$ is included via the
Peierls phase factor $a_{\vec{i},\vec{j}}=\frac{2\mathrm{\pi}
e}{h}\int_\vec{j}^\vec{i}\vec{A}\cdot\mathrm{d}\vec{l}$, where
$\vec{A}=(0,-Bx,0)$ is the vector potential in the Landau gauge. We note
that usually a perpendicular magnetic field, even as large as $60 T$, has little
effect on the DDW gap,~\cite{Nguyen:2002} except close to the doping at which it collapses, where field induced order may be important.

We have seen previously~\cite{Jia:2009} that the effect of long-ranged correlated disorder is qualitatively similar to white noise insofar as the  QOs are concerned. The effect of the nature of disorder on the spectral  function of angle resolved photoemission spectroscopy (ARPES) was found to be far more important. The reason is that the coherence factors of the ARPES spectral function are sensitive to the nature of the disorder because they play a role similar to Wannier functions.
In contrast, the QOs are damped by the Dingle factor, which is parametrized by a single
lifetime and disorder enters in an averaged sense. 

Thus, it is sufficient to consider on-site disorder.
The on-site energy is $\delta$-correlated white noise defined by the disorder average $\overline{\epsilon_\vec{i}}=0$ and $\overline{\epsilon_\vec{i}\epsilon_\vec{j}}=V_{0}^{2}\delta_{\vec{i},\vec{j}}$. For an explicit calculation we need to choose the band structure parameters, $W_{0}$, $V_{S}$, and the disorder magnitude $V_{0}$.
When considering the magnitude of disorder one should keep in mind that the full band width is $8 t$.  The magnetic field
ranges roughly between  $30 T $ and $64 T$, representative
of the experiments in NCCO. The  magnetic length is $l_B=\sqrt{\hbar/eB}$, which for
 $B=30 T$ is  approximately $ 12 a$, where the lattice constant $a$ is equal to
$3.95\textrm{\AA}$.

The effect of potential scattering that modulates charge density is indirect on two-fold commensurate SDW or DDW order parameter,~\cite{Ghosal:2004} mainly because SDW is modulation of spin and DDW that of charge current.  Thus, the robustness of these order parameters with respect to disorder protects the corresponding quasiparticle excitations insofar as quantum oscillations are concerned, as seen below  in our exact numerical calculations. Thus we did not find it important to study this problem self consistently.

\section{Transfer matrix method}

The transfer matrix method and the calculation of the Lyapunov sketched  elsewhere~\cite{Jia:2009} is fully described here for the case of singlet DDW; for SDW the generalization is straightforward, where the diagonal term must be modified because of $V_S$, and the term $W_0$ will be absent.
Consider a quasi-1D system, $L\gg M$, with a periodic boundary
condition along y-direction. Let $\Psi_n = (\psi_{n,1},\psi_{n,2},
\ldots, \psi_{n,M})^T$ be the amplitudes on the slice $n$ for an
eigenstate with a given energy, then the amplitudes on three
successive slices satisfy the relation
\begin{equation}
    \label{Eq:transfermatrix}
    \left[
      \begin{array}{c}
        \Psi_{n+1} \\
        \Psi_{n} \\
      \end{array}
    \right] = \left[
                \begin{array}{cc}
                  T_n^{-1}A_{n} & -T_n^{-1}B_{n}\\
                  1 & 0 \\
                \end{array}
              \right]\left[
                       \begin{array}{c}
                         \Psi_n \\
                         \Psi_{n-1} \\
                       \end{array}
                     \right]={\mathbf T}_{n}\left[
                       \begin{array}{c}
                         \Psi_n \\
                         \Psi_{n-1} \\
                       \end{array}\right]
\end{equation}
where $T_n$, $A_{n}$, $B_{n}$ are  $M\times M$ matrices.
The non-zero matrix elements of the matrix $A_n$ are
\begin{equation}
\begin{split}
   (A_n)_{m,m}&=\epsilon_{n,m}-\mu,\\
   (A_n)_{m,m+1}&=\left[-t+\frac{\mi W_0}{4}(-1)^{m+n}\right]\me^{-\mi n\phi},\\
   (A_n)_{m,m-1}&=\left[-t+\frac{\mi W_0}{4}(-1)^{m+n}\right]\me^{\mi n\phi}.
 \end{split}
\end{equation}
where $\phi = 2\pi Ba^2e/h$ is a constant. For the matrix $B_n$:
\begin{equation}
\begin{split}
   (B_n)_{m,m}&=-\left[-t-\frac{\mi W_0}{4}(-1)^{m+n}\right],\\
   (B_n)_{m,m+1}&=-t'\me^{\mi(-n+\frac{1}{2})\phi},\\
   (B_n)_{m,m-1}&=-t'\me^{\mi(n-\frac{1}{2})\phi},
\end{split}
\end{equation}
For the matrix $T_n$, we note that $T_n = B^{\dag}_{n+1}$.

The $2M$ Lyapunov exponents, $\gamma_{i}$, of $\lim_{N\to \infty}({\cal T}_{N}{\cal T}_{N}^{\dagger})^{1/2N}$, where $ {\cal T}_{N}=\prod_{j=1}^{j=N}{\mathbf T}_{j}$, are defined by the corresponding eigenvalues $\lambda_{i}=e^{\gamma_{i}}$. 
 All Lyapunov exponents
$\gamma_1>\gamma_2>\ldots>\gamma_{2M}$, are computed by a procedure given in Ref.~\onlinecite{Kramer:1996}. 
The modification here is that this matrix is not symplectic. Therefore all $2M$ eigenvalues have to be computed. The remarkable fact, however, is that except for a small fraction, consisting of larger eigenvalues, the rest do come in pairs $(\lambda, 1/\lambda)$, as for the symplectic case, within numerical accuracy. We have no analytical proof of this curious fact. Clearly, larger eigenvalues contribute insignificantly to the more general formula for the conductance:~\cite{Pichard:1986}
\begin{equation}
\sigma(B) = \frac{e^{2}}{h} \text{Tr}\sum_{j=1}^{2M}\frac{2}{({\cal T}_{N}{\cal T}_{N}^{\dagger})+({\cal T}_{N}{\cal T}_{N}^{\dagger})^{-1}+2}.
\end{equation}
When the eigenvalues do come in pairs,  the conductance formula simplifies to the more common  Landauer formula:~\cite{Fisher:1981}
\begin{equation}\label{Eq:landauer}
    \sigma_{xx}(B)=\frac{e^2}{h}\sum_{i=1}^{M} \frac{1}{\cosh^2
    (M\gamma_i)}.
\end{equation}

The transfer matrix method is a very powerful method and the results obtained are rigorous compared to {\em ad hoc} broadening of the Landau levels, which also require more adjustable parameters to explain the experiments. Once the distribution of disorder is specified there are no further approximations.  We note that the values of $M$ were chosen to be much larger than our previous work,~\cite{Jia:2009} at least 128 (that is $128 \; a $ in physical units) and sometimes as large as 512. The length of the strip  $L$ is varied between $10^{5}$ and $10^{6}$. This easily led to an accuracy better than 5\% for the smallest Lyapunov exponent, $\gamma_{i}$, in all cases.   

We have calculated the $ab$-plane conductance, but the measured $c$-axis resistance, $R_{c}$, is precisely related to it, at least  as far as  the oscillatory part is concerned. This can be seen from the arguments in Ref.~\onlinecite{Kumar:1992}. Although the details can be improved,  the crux of the argument is that the planar density of states enters $R_{c}$: the quasiparticle scatters many times in the plane while performing cyclotron motion before hopping from plane to plane (measured $ab$-plane resistivity is of the order $10\mu\Omega$-cm as compared $\Omega$-cm for the $c$-axis resistivity even at optimum doping~\cite{Armitage:2009}). It is worth noting that oscillations of $R_{c}$ also precisely follows the oscillations of the magnetization in overdoped $\mathrm{Tl_{2}Ba_{2}CuO_{6+\delta}}$.~\cite{vignolle:2008}

\section{Results}
There are clues in the experiments~\cite{Helm:2009} that disorder is very important.  For 15 and 16\% doping the slow oscillations in experiments, of frequency $290-280 T$, are not observed until the field reaches above $ 30 T$, which is much greater than $H_{c2} < 10 T$. For 17\% doping the onset of fast oscillations at a frequency of  $10, 700 T$ are strikingly not observable until the field reaches $60 T$. The estimated scattering time from the Dingle factor at even optimal doping and at $4 K$ is quite short.

\begin{figure}
\begin{center}
\includegraphics[scale=0.4]{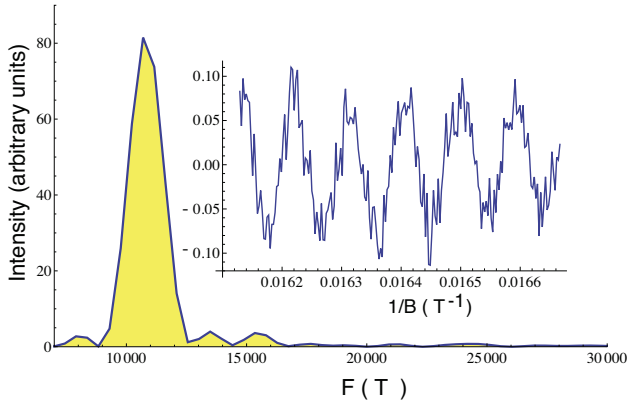}
\caption{(Color online)The main plot shows the Fourier transform of the field sweep shown in the inset. The peak is at $10,695 T$. The inset is a smooth background subtracted Shubnikov-de Haas oscillations, as calculated from the Landauer formula for 17\% doping as a function of $1/B$. The disorder parameter is $V_{0}=0.7 t$.The band structure parameters are given in Table~\ref{table1}.}
\label{fig:NCCO-17-7}
\end{center}
\end{figure}

\begin{figure}
\begin{center}
\includegraphics[scale=0.4]{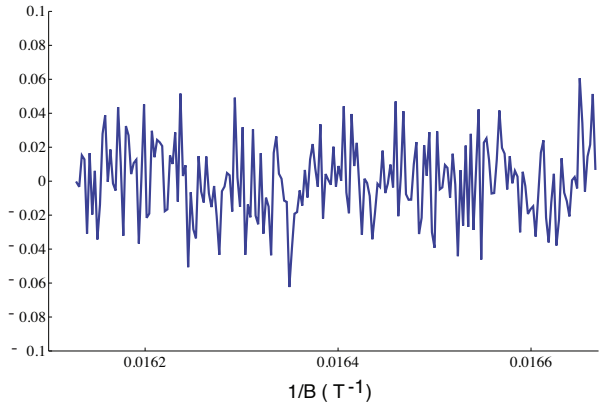}
\caption{(Color online)The same parameters as in Fig.~\ref{fig:NCCO-17-7} but $V_{0}=0.8 t$. The background subtracted conductance is simply noise to an excellent approximation.}
\label{fig:NCCO-17-8}
\end{center}
\end{figure}

For 17\% doping corresponding to $\mu = -0.322 t$ and the band structure given in  Table~\ref{table1}, a slight change in disorder from $V_{0}=0.7 t$ to $V_{0}=0.8 t$ makes the difference between a clear observation of a peak to simply noise within the field sweep between $60-62 T$, as shown in Fig.~\ref{fig:NCCO-17-7} and Fig.~\ref{fig:NCCO-17-8}. Since in this case $W_{0}=V_{S}=0$, there is little else to blame for the disappearance of the oscillations for fields roughly below $60 T$. The results are essentially identical for small values of $W_{0}$, such as $0.025 t$.

\begin{figure}
\begin{center}
\includegraphics[scale=0.4]{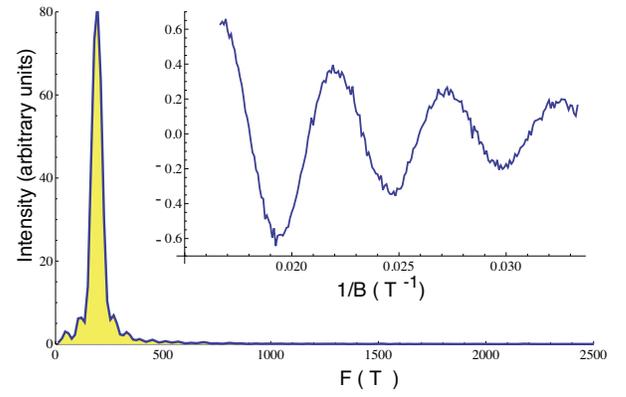}
\caption{(Color online)The same plot as in Fig.~\ref{fig:NCCO-17-7}, except for  15\% doping and DDW order. The parameters are given in Table~\ref{table1}.}
\label{fig:NCCO-15}
\end{center}
\end{figure}

\begin{figure}
\begin{center}
\includegraphics[scale=0.4]{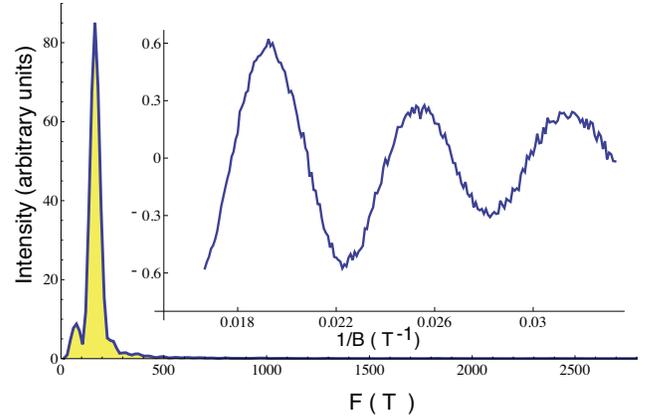}
\caption{(Color online)The same plot as in Fig.~\ref{fig:NCCO-17-7}, except for  16\% doping and DDW order. The parameters are given in Table~\ref{table1}.}
\label{fig:NCCO-16}
\end{center}
\end{figure}

\begin{figure}
\begin{center}
\includegraphics[scale=0.4]{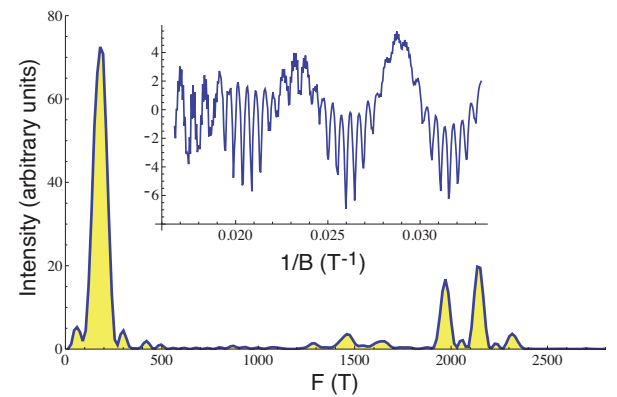}
\caption{(Color online) The same plot as in Fig~\ref{fig:NCCO-15}, except  that $V_{0}=0.2 t$ instead of $0.8t$. There is now a fragmented electron pocket centered around $2100 T$ and the main peak is at $183 T$. The rest of the parameters are given in Table~\ref{table1}.}
\label{fig:NCCO-15-4}
\end{center}
\end{figure}

For 15\% and 16\% dopings we chose $V_{0}$ to simulate the fact that oscillations seem to disappear below $30 T$. The field sweep was between $30-60 T$.   The results for DDW order  are shown in Fig.~\ref{fig:NCCO-15} and Fig.~\ref{fig:NCCO-16}.
 The most remarkable feature of these figures  is that disorder has completely wiped out the large electron pocket leaving the small hole pocket visible. To emphasize this point we also plot the results for 15\% doping but with much smaller disorder $V_{0}=0.2 t$; see Fig.~\ref{fig:NCCO-15-4}. Now we can see the fragmented remnants of the electron pocket. With further lowering of disorder, the full electron pocket becomes visible. It is clear that disorder has a significantly stronger effect on the electron pockets than on the hole pockets. This, as we noted earlier, is largely due to higher density of states around the antinodal points, which significantly accentuates the effect of disorder.~\cite{Jia:2009}
\begin{figure}
\begin{center}
\includegraphics[scale=0.4]{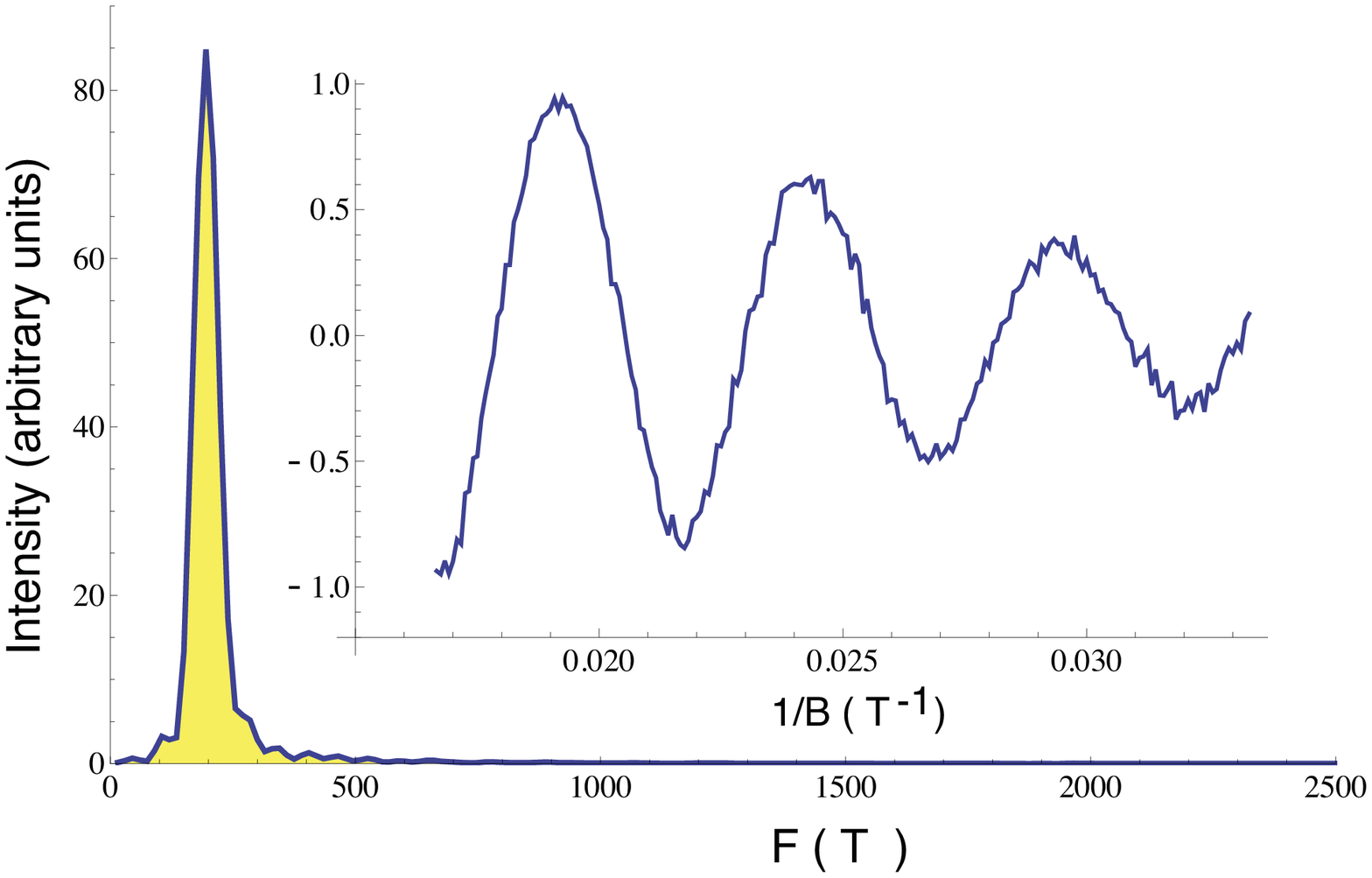}
\caption{(Color online) The same plot as in Fig.~\ref{fig:NCCO-15} for 15\% doping but using SDW order.  The main peak is at $195 T$.   The rest of the parameters are given in Table~\ref{table1}.}
\label{fig:SDW-15}
\end{center}
\end{figure}
We have done parallel calculations with SDW order as well. The results are essentially identical. They are shown again  for 15 and 16\% doping in Fig.~\ref{fig:SDW-15} and Fig.~\ref{fig:SDW-16}. We have kept all parameters fixed, while adjusting the the SDW gap to achieve as best an approximation to experiments as possible.
\begin{figure}
\begin{center}
\includegraphics[scale=0.4]{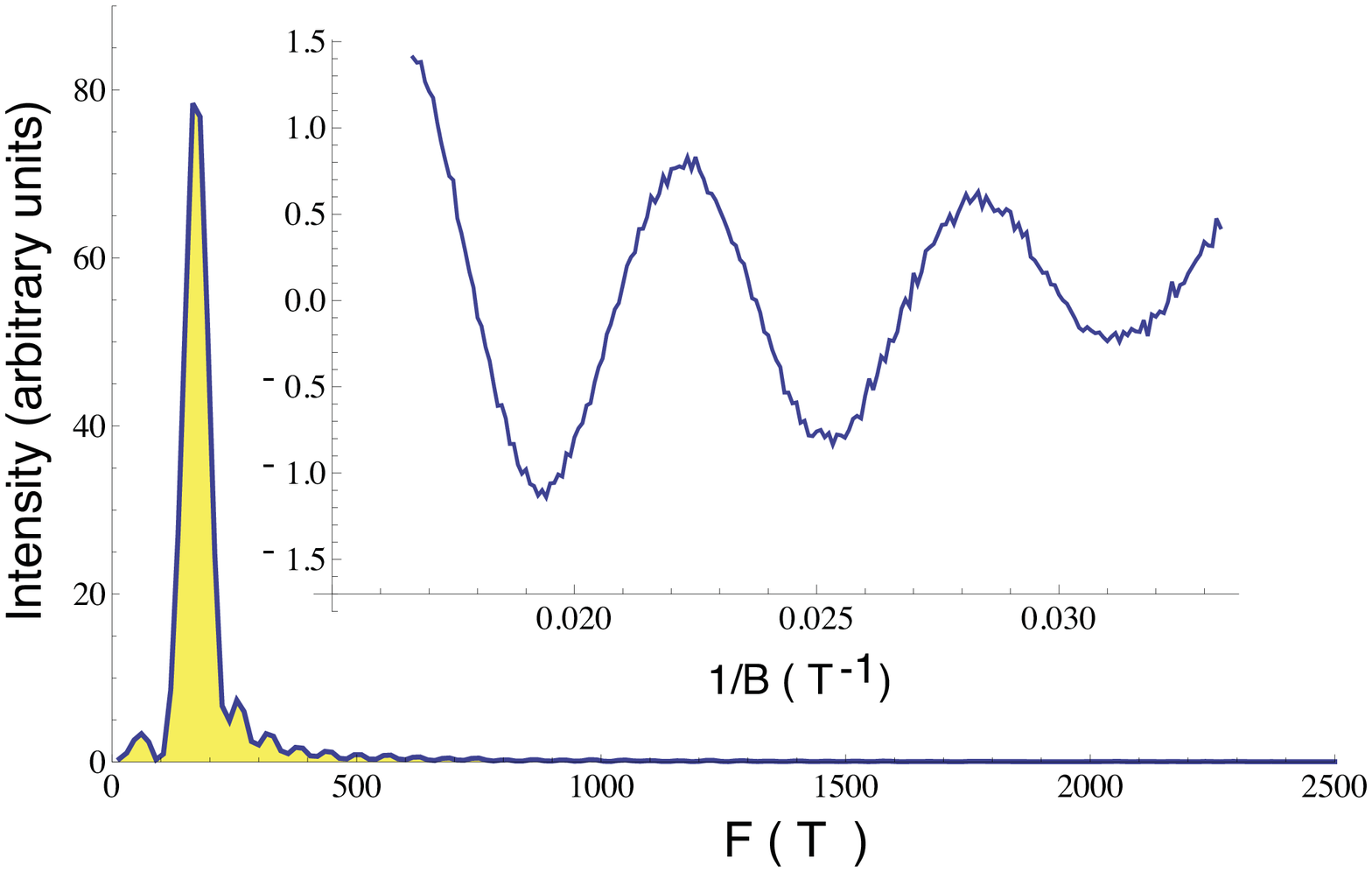}
\caption{(Color online) The same plot as in Fig.~\ref{fig:SDW-15}, except for 16\% doping and  using SDW order.  The main peak is at $173 T$.   The rest of the parameters are given in Table~\ref{table1}.}
\label{fig:SDW-16}
\end{center}
\end{figure}

It is important to summarize our results in the context of experimental observations. First, we were able to show that the electron pocket frequencies are strikingly  absent because of disorder and the slow frequencies corresponding to the hole pocket for 15\% and 16\% doping damp out below about 30 T, even though $H_{c2}$ is less than 10 T. Similarly, that the high frequency oscillations at 17\% doping do not arise until about 60 T has a natural explanation in terms of disorder, although in this case some magnetic breakdown effect, which was not explored, can be expected. This requires both further experimental and theoretical investigations. The calculated frequency of the high frequency oscillations, $10,695$ T is remarkably close to experimental value of $10,700 \pm 400$ T. As to the magnitude of the slow oscillations, the calculated values are given in Table~\ref{table1}, which are reasonable in both magnitude and trend when compared to experiments. The small discrepancies in the magnitude of $F$  are due to our neglect of $t''$ in the band structure. This can be, and was, checked by checking the pure case, that is, without disorder.

\section{Conclusions}

In the absence of disorder or thermal broadening, the oscillation waveforms are never sinusoidal in two dimensions and contain many Fourier harmonics. At zero temperature moderate disorder converts  the oscillations to sinusoidal waveform with rapidly decreasing amplitudes of the harmonics. Further increase of disorder ultimately destroys the amplitudes altogether. Many experiments exhibit roughly sinusoidal waveform at even ultra low temperatures, implying  that disorder is important. The remarkably small electronic dispersion in the direction perpendicular to the CuO-planes cannot alone account for the waveform.

For NCCO it is no longer a mystery as to why the frequency corresponding to the  larger electron pocket is not observed. As we have shown, disorder is the culprit. Neither is the comparison with ARPES controversial,~\cite{Armitage:2009} as in the case of YBCO, since there is good evidence of Fermi surface crossing in the direction $(\pi,0)\to (\pi,\pi)$, which is a signature of the electron pocket. The crossing along $(\pi,\pi)\to (0,0)$ can be easily construed as an evidence of a small hole pocket for which half of it is made invisible both from the coherence factors and disorder effects.~\cite{Jia:2009} For electron doped materials, such as NCCO and PCCO, it is known~\cite{Armitage:2009}  that the Hall coefficient changes sign around 17\% doping and therefore the picture of reconnection of the Fermi pockets  is entirely plausible, with some likely magnetic breakdown effects. The real question is what is the evidence of SDW or DDW in the relevant doping range between 15\% and 17\%. From neutron measurements we know that there is no long range SDW order for doping above 13.4\%.~\cite{Motoyama:2007} We cannot rule out field induced SDW at about $30 T$. For DDW, there are no corresponding neutron  measurements to observe its existence. Given that DDW is  considerably more hidden~\cite{Nayak:2000,Chakravarty:2001} from common experiments, it is more challenging to establish it directly. NMR experiments in high fields for suitable nuclei can shed light on this question. The unavoidable logical conclusion from the QO measurements is that a density wave that breaks translational symmetry must be present. We suggest that motivated future experiments will be necessary to reach a definitive conclusion. Finally, at the level of mean field theory we have been unable to decide between SDW and singlet DDW. At the moment the best recourse is to experimentally look for spin zeros in the amplitude of quantum oscillations in a tilted magnetic field. A theoretical discussion of this phenomenon that can potentially shed light between a triplet order parameter (SDW) and a singlet order parameter, the singlet DDW discussed here, was provided recently.~\cite{Garcia:2010} So far experiments are in conflict with each other in YBCO: one group suggests a triplet order parameter~\cite{Sebastian:2009,Sebastian:2010} and the other a singlet order parameter.~\cite{Ramshaw:2010}

It is unquestionable that the QO experiments are likely to change the widespread views in the field of high temperature superconductivity. Although the measurements in YBCO are not fully explained, the measurements in NCCO appear to have a clear and simple explanation, as shown here. However, given the similarity of the phenomenon in both hole and  electron doped cuprates, it is likely that the quantum oscillations have the same origin.

\acknowledgments
This work is supported by NSF under the Grant DMR-0705092. All calculations were performed at Hoffman 2 Cluster at UCLA. We thank E. Abrahams  for a critical reading of the manuscript and N. P. Armitage for comments.


\begin{thebibliography}{32}
\expandafter\ifx\csname natexlab\endcsname\relax\def\natexlab#1{#1}\fi
\expandafter\ifx\csname bibnamefont\endcsname\relax
  \def\bibnamefont#1{#1}\fi
\expandafter\ifx\csname bibfnamefont\endcsname\relax
  \def\bibfnamefont#1{#1}\fi
\expandafter\ifx\csname citenamefont\endcsname\relax
  \def\citenamefont#1{#1}\fi
\expandafter\ifx\csname url\endcsname\relax
  \def\url#1{\texttt{#1}}\fi
\expandafter\ifx\csname urlprefix\endcsname\relax\def\urlprefix{URL }\fi
\providecommand{\bibinfo}[2]{#2}
\providecommand{\eprint}[2][]{\url{#2}}

\bibitem[{\citenamefont{Doiron-Leyraud
  et~al.}(2007)\citenamefont{Doiron-Leyraud, Proust, LeBoeuf, Levallois,
  Bonnemaison, Liang, Bonn, Hardy, and Taillefer}}]{Doiron-Leyraud:2007}
\bibinfo{author}{\bibfnamefont{N.}~\bibnamefont{Doiron-Leyraud}},
  \bibinfo{author}{\bibfnamefont{C.}~\bibnamefont{Proust}},
  \bibinfo{author}{\bibfnamefont{D.}~\bibnamefont{LeBoeuf}},
  \bibinfo{author}{\bibfnamefont{J.}~\bibnamefont{Levallois}},
  \bibinfo{author}{\bibfnamefont{J.-B.} \bibnamefont{Bonnemaison}},
  \bibinfo{author}{\bibfnamefont{R.}~\bibnamefont{Liang}},
  \bibinfo{author}{\bibfnamefont{D.~A.} \bibnamefont{Bonn}},
  \bibinfo{author}{\bibfnamefont{W.~N.} \bibnamefont{Hardy}}, \bibnamefont{and}
  \bibinfo{author}{\bibfnamefont{L.}~\bibnamefont{Taillefer}},
  \bibinfo{journal}{Nature} \textbf{\bibinfo{volume}{447}},
  \bibinfo{pages}{565} (\bibinfo{year}{2007}).

\bibitem[{\citenamefont{Bangura et~al.}(2008)\citenamefont{Bangura, Fletcher,
  Carrington, Levallois, Nardone, Vignolle, Heard, Doiron-Leyraud, LeBoeuf,
  Taillefer et~al.}}]{Bangura:2008}
\bibinfo{author}{\bibfnamefont{A.~F.} \bibnamefont{Bangura}},
  \bibinfo{author}{\bibfnamefont{J.~D.} \bibnamefont{Fletcher}},
  \bibinfo{author}{\bibfnamefont{A.}~\bibnamefont{Carrington}},
  \bibinfo{author}{\bibfnamefont{J.}~\bibnamefont{Levallois}},
  \bibinfo{author}{\bibfnamefont{M.}~\bibnamefont{Nardone}},
  \bibinfo{author}{\bibfnamefont{B.}~\bibnamefont{Vignolle}},
  \bibinfo{author}{\bibfnamefont{P.~J.} \bibnamefont{Heard}},
  \bibinfo{author}{\bibfnamefont{N.}~\bibnamefont{Doiron-Leyraud}},
  \bibinfo{author}{\bibfnamefont{D.}~\bibnamefont{LeBoeuf}},
  \bibinfo{author}{\bibfnamefont{L.}~\bibnamefont{Taillefer}},
  \bibnamefont{et~al.}, \bibinfo{journal}{Phys. Rev. Lett.}
  \textbf{\bibinfo{volume}{100}}, \bibinfo{pages}{047004}
  (\bibinfo{year}{2008}).

\bibitem[{\citenamefont{LeBoeuf et~al.}(2007)\citenamefont{LeBoeuf,
  Doiron-Leyraud, Levallois, Daou, Bonnemaison, Hussey, Balicas, Ramshaw,
  Liang, Bonn et~al.}}]{LeBoeuf:2007}
\bibinfo{author}{\bibfnamefont{D.}~\bibnamefont{LeBoeuf}},
  \bibinfo{author}{\bibfnamefont{N.}~\bibnamefont{Doiron-Leyraud}},
  \bibinfo{author}{\bibfnamefont{J.}~\bibnamefont{Levallois}},
  \bibinfo{author}{\bibfnamefont{R.}~\bibnamefont{Daou}},
  \bibinfo{author}{\bibfnamefont{J.~B.} \bibnamefont{Bonnemaison}},
  \bibinfo{author}{\bibfnamefont{N.~E.} \bibnamefont{Hussey}},
  \bibinfo{author}{\bibfnamefont{L.}~\bibnamefont{Balicas}},
  \bibinfo{author}{\bibfnamefont{B.~J.} \bibnamefont{Ramshaw}},
  \bibinfo{author}{\bibfnamefont{R.}~\bibnamefont{Liang}},
  \bibinfo{author}{\bibfnamefont{D.~A.} \bibnamefont{Bonn}},
  \bibnamefont{et~al.}, \bibinfo{journal}{Nature}
  \textbf{\bibinfo{volume}{450}}, \bibinfo{pages}{533} (\bibinfo{year}{2007}).

\bibitem[{\citenamefont{Jaudet et~al.}(2008)\citenamefont{Jaudet, Vignolles,
  Audouard, Levallois, LeBoeuf, Doiron-Leyraud, Vignolle, Nardone, Zitouni,
  Liang et~al.}}]{Jaudet:2008}
\bibinfo{author}{\bibfnamefont{C.}~\bibnamefont{Jaudet}},
  \bibinfo{author}{\bibfnamefont{D.}~\bibnamefont{Vignolles}},
  \bibinfo{author}{\bibfnamefont{A.}~\bibnamefont{Audouard}},
  \bibinfo{author}{\bibfnamefont{J.}~\bibnamefont{Levallois}},
  \bibinfo{author}{\bibfnamefont{D.}~\bibnamefont{LeBoeuf}},
  \bibinfo{author}{\bibfnamefont{N.}~\bibnamefont{Doiron-Leyraud}},
  \bibinfo{author}{\bibfnamefont{B.}~\bibnamefont{Vignolle}},
  \bibinfo{author}{\bibfnamefont{M.}~\bibnamefont{Nardone}},
  \bibinfo{author}{\bibfnamefont{A.}~\bibnamefont{Zitouni}},
  \bibinfo{author}{\bibfnamefont{R.}~\bibnamefont{Liang}},
  \bibnamefont{et~al.}, \bibinfo{journal}{Phys. Rev. Lett.}
  \textbf{\bibinfo{volume}{100}}, \bibinfo{pages}{187005}
  (\bibinfo{year}{2008}).

\bibitem[{\citenamefont{Yelland et~al.}(2008)\citenamefont{Yelland, Singleton,
  Mielke, Harrison, Balakirev, Dabrowski, and Cooper}}]{Yelland:2008}
\bibinfo{author}{\bibfnamefont{E.~A.} \bibnamefont{Yelland}},
  \bibinfo{author}{\bibfnamefont{J.}~\bibnamefont{Singleton}},
  \bibinfo{author}{\bibfnamefont{C.~H.} \bibnamefont{Mielke}},
  \bibinfo{author}{\bibfnamefont{N.}~\bibnamefont{Harrison}},
  \bibinfo{author}{\bibfnamefont{F.~F.} \bibnamefont{Balakirev}},
  \bibinfo{author}{\bibfnamefont{B.}~\bibnamefont{Dabrowski}},
  \bibnamefont{and} \bibinfo{author}{\bibfnamefont{J.~R.}
  \bibnamefont{Cooper}}, \bibinfo{journal}{Phys. Rev. Lett.}
  \textbf{\bibinfo{volume}{100}}, \bibinfo{pages}{047003}
  (\bibinfo{year}{2008}).

\bibitem[{\citenamefont{Sebastian et~al.}(2008)\citenamefont{Sebastian,
  Harrison, Palm, Murphy, Mielke, Liang, Bonn, Hardy, and
  Lonzarich}}]{Sebastian:2008}
\bibinfo{author}{\bibfnamefont{S.~E.} \bibnamefont{Sebastian}},
  \bibinfo{author}{\bibfnamefont{N.}~\bibnamefont{Harrison}},
  \bibinfo{author}{\bibfnamefont{E.}~\bibnamefont{Palm}},
  \bibinfo{author}{\bibfnamefont{T.~P.} \bibnamefont{Murphy}},
  \bibinfo{author}{\bibfnamefont{C.~H.} \bibnamefont{Mielke}},
  \bibinfo{author}{\bibfnamefont{R.}~\bibnamefont{Liang}},
  \bibinfo{author}{\bibfnamefont{D.~A.} \bibnamefont{Bonn}},
  \bibinfo{author}{\bibfnamefont{W.~N.} \bibnamefont{Hardy}}, \bibnamefont{and}
  \bibinfo{author}{\bibfnamefont{G.~G.} \bibnamefont{Lonzarich}},
  \bibinfo{journal}{Nature} \textbf{\bibinfo{volume}{454}},
  \bibinfo{pages}{200} (\bibinfo{year}{2008}).

\bibitem[{\citenamefont{Audouard et~al.}(2009)\citenamefont{Audouard, Jaudet,
  Vignolles, Liang, Bonn, Hardy, Taillefer, and Proust}}]{Audouard:2009}
\bibinfo{author}{\bibfnamefont{A.}~\bibnamefont{Audouard}},
  \bibinfo{author}{\bibfnamefont{C.}~\bibnamefont{Jaudet}},
  \bibinfo{author}{\bibfnamefont{D.}~\bibnamefont{Vignolles}},
  \bibinfo{author}{\bibfnamefont{R.}~\bibnamefont{Liang}},
  \bibinfo{author}{\bibfnamefont{D.}~\bibnamefont{Bonn}},
  \bibinfo{author}{\bibfnamefont{W.}~\bibnamefont{Hardy}},
  \bibinfo{author}{\bibfnamefont{L.}~\bibnamefont{Taillefer}},
  \bibnamefont{and} \bibinfo{author}{\bibfnamefont{C.}~\bibnamefont{Proust}},
  \bibinfo{journal}{Phys. Rev. Lett.} \textbf{\bibinfo{volume}{103}},
  \bibinfo{pages}{157003} (\bibinfo{year}{2009}).

\bibitem[{\citenamefont{Singleton et~al.}(2010)\citenamefont{Singleton,
  De~La~Cruz, McDonald, Li, Altarawneh, Goddard, Franke, Rickel, Mielke, Yao
  et~al.}}]{Singleton:2009}
\bibinfo{author}{\bibfnamefont{J.}~\bibnamefont{Singleton}},
  \bibinfo{author}{\bibfnamefont{C.}~\bibnamefont{De~La~Cruz}},
  \bibinfo{author}{\bibfnamefont{R.~D.} \bibnamefont{McDonald}},
  \bibinfo{author}{\bibfnamefont{S.}~\bibnamefont{Li}},
  \bibinfo{author}{\bibfnamefont{M.}~\bibnamefont{Altarawneh}},
  \bibinfo{author}{\bibfnamefont{P.}~\bibnamefont{Goddard}},
  \bibinfo{author}{\bibfnamefont{I.}~\bibnamefont{Franke}},
  \bibinfo{author}{\bibfnamefont{D.}~\bibnamefont{Rickel}},
  \bibinfo{author}{\bibfnamefont{C.~H.} \bibnamefont{Mielke}},
  \bibinfo{author}{\bibfnamefont{X.}~\bibnamefont{Yao}}, \bibnamefont{et~al.},
  \bibinfo{journal}{Phys. Rev. Lett.} p. \bibinfo{pages}{086403}
  (\bibinfo{year}{2010}).

\bibitem[{\citenamefont{Chakravarty}(2008)}]{Chakravarty:2008}
\bibinfo{author}{\bibfnamefont{S.}~\bibnamefont{Chakravarty}},
  \bibinfo{journal}{Science} \textbf{\bibinfo{volume}{319}},
  \bibinfo{pages}{735} (\bibinfo{year}{2008}).

\bibitem[{\citenamefont{Vignolle et~al.}(2008)\citenamefont{Vignolle,
  Carrington, Cooper, French, Mackenzie, Jaudet, Vignolles, Proust, and
  Hussey}}]{vignolle:2008}
\bibinfo{author}{\bibfnamefont{B.}~\bibnamefont{Vignolle}},
  \bibinfo{author}{\bibfnamefont{A.}~\bibnamefont{Carrington}},
  \bibinfo{author}{\bibfnamefont{R.~A.} \bibnamefont{Cooper}},
  \bibinfo{author}{\bibfnamefont{M.~M.~J.} \bibnamefont{French}},
  \bibinfo{author}{\bibfnamefont{A.~P.} \bibnamefont{Mackenzie}},
  \bibinfo{author}{\bibfnamefont{C.}~\bibnamefont{Jaudet}},
  \bibinfo{author}{\bibfnamefont{D.}~\bibnamefont{Vignolles}},
  \bibinfo{author}{\bibfnamefont{C.}~\bibnamefont{Proust}}, \bibnamefont{and}
  \bibinfo{author}{\bibfnamefont{N.~E.} \bibnamefont{Hussey}},
  \bibinfo{journal}{Nature} \textbf{\bibinfo{volume}{455}},
  \bibinfo{pages}{952} (\bibinfo{year}{2008}).

\bibitem[{\citenamefont{Helm et~al.}(2009)\citenamefont{Helm, Kartsovnik,
  Bartkowiak, Bittner, Lambacher, Erb, Wosnitza, and Gross}}]{Helm:2009}
\bibinfo{author}{\bibfnamefont{T.}~\bibnamefont{Helm}},
  \bibinfo{author}{\bibfnamefont{M.~V.} \bibnamefont{Kartsovnik}},
  \bibinfo{author}{\bibfnamefont{M.}~\bibnamefont{Bartkowiak}},
  \bibinfo{author}{\bibfnamefont{N.}~\bibnamefont{Bittner}},
  \bibinfo{author}{\bibfnamefont{M.}~\bibnamefont{Lambacher}},
  \bibinfo{author}{\bibfnamefont{A.}~\bibnamefont{Erb}},
  \bibinfo{author}{\bibfnamefont{J.}~\bibnamefont{Wosnitza}}, \bibnamefont{and}
  \bibinfo{author}{\bibfnamefont{R.}~\bibnamefont{Gross}},
  \bibinfo{journal}{Phys. Rev. Lett.} \textbf{\bibinfo{volume}{103}},
  \bibinfo{pages}{157002} (\bibinfo{year}{2009}).

\bibitem[{\citenamefont{Podolsky and Kee}(2008)}]{Podolsky:2008}
\bibinfo{author}{\bibfnamefont{D.}~\bibnamefont{Podolsky}} \bibnamefont{and}
  \bibinfo{author}{\bibfnamefont{H.-Y.} \bibnamefont{Kee}},
  \bibinfo{journal}{Phys. Rev. B} \textbf{\bibinfo{volume}{78}},
  \bibinfo{pages}{224516} (\bibinfo{year}{2008}).

\bibitem[{\citenamefont{Millis and Norman}(2007)}]{Millis:2007}
\bibinfo{author}{\bibfnamefont{A.~J.} \bibnamefont{Millis}} \bibnamefont{and}
  \bibinfo{author}{\bibfnamefont{M.~R.} \bibnamefont{Norman}},
  \bibinfo{journal}{Phys. Rev. B} \textbf{\bibinfo{volume}{76}},
  \bibinfo{pages}{220503} (\bibinfo{year}{2007}).

\bibitem[{\citenamefont{Armitage et~al.}(2009)\citenamefont{Armitage, Fournier,
  and Green}}]{Armitage:2009}
\bibinfo{author}{\bibfnamefont{N.~P.} \bibnamefont{Armitage}},
  \bibinfo{author}{\bibfnamefont{P.}~\bibnamefont{Fournier}}, \bibnamefont{and}
  \bibinfo{author}{\bibfnamefont{R.~L.} \bibnamefont{Green}},
  \bibinfo{journal}{arXiv:0906.2931}  (\bibinfo{year}{2009}).

\bibitem[{\citenamefont{Chakravarty et~al.}(2001)\citenamefont{Chakravarty,
  Laughlin, Morr, and Nayak}}]{Chakravarty:2001}
\bibinfo{author}{\bibfnamefont{S.}~\bibnamefont{Chakravarty}},
  \bibinfo{author}{\bibfnamefont{R.~B.} \bibnamefont{Laughlin}},
  \bibinfo{author}{\bibfnamefont{D.~K.} \bibnamefont{Morr}}, \bibnamefont{and}
  \bibinfo{author}{\bibfnamefont{C.}~\bibnamefont{Nayak}},
  \bibinfo{journal}{Phys. Rev. B} \textbf{\bibinfo{volume}{63}},
  \bibinfo{pages}{094503} (\bibinfo{year}{2001}).

\bibitem[{\citenamefont{Rourke et~al.}(2009)\citenamefont{Rourke, Bangura,
  Proust, Levallois, Doiron-Leyraud, LeBoeuf, Taillefer, Adachi, Sutherland,
  and Hussey}}]{Rourke:2009}
\bibinfo{author}{\bibfnamefont{P.~M.~C.} \bibnamefont{Rourke}},
  \bibinfo{author}{\bibfnamefont{A.~F.} \bibnamefont{Bangura}},
  \bibinfo{author}{\bibfnamefont{C.}~\bibnamefont{Proust}},
  \bibinfo{author}{\bibfnamefont{J.}~\bibnamefont{Levallois}},
  \bibinfo{author}{\bibfnamefont{N.}~\bibnamefont{Doiron-Leyraud}},
  \bibinfo{author}{\bibfnamefont{D.}~\bibnamefont{LeBoeuf}},
  \bibinfo{author}{\bibfnamefont{L.}~\bibnamefont{Taillefer}},
  \bibinfo{author}{\bibfnamefont{S.}~\bibnamefont{Adachi}},
  \bibinfo{author}{\bibfnamefont{M.~L.} \bibnamefont{Sutherland}},
  \bibnamefont{and} \bibinfo{author}{\bibfnamefont{N.~E.}
  \bibnamefont{Hussey}}, \bibinfo{journal}{http://arxiv.org/abs/0912.0175}
  (\bibinfo{year}{2009}).

\bibitem[{\citenamefont{Kusko et~al.}(2002)\citenamefont{Kusko, Markiewicz,
  Lindroos, and Bansil}}]{Kusko:2002}
\bibinfo{author}{\bibfnamefont{C.}~\bibnamefont{Kusko}},
  \bibinfo{author}{\bibfnamefont{R.~S.} \bibnamefont{Markiewicz}},
  \bibinfo{author}{\bibfnamefont{M.}~\bibnamefont{Lindroos}}, \bibnamefont{and}
  \bibinfo{author}{\bibfnamefont{A.}~\bibnamefont{Bansil}},
  \bibinfo{journal}{Phys. Rev. B} \textbf{\bibinfo{volume}{66}},
  \bibinfo{pages}{140513} (\bibinfo{year}{2002}).

\bibitem[{\citenamefont{Chakravarty and Kee}(2008)}]{Chakravarty:2008b}
\bibinfo{author}{\bibfnamefont{S.}~\bibnamefont{Chakravarty}} \bibnamefont{and}
  \bibinfo{author}{\bibfnamefont{H.-Y.} \bibnamefont{Kee}},
  \bibinfo{journal}{Proc. Natl. Acad. Sci. USA} \textbf{\bibinfo{volume}{105}},
  \bibinfo{pages}{8835} (\bibinfo{year}{2008}).

\bibitem[{\citenamefont{Jia et~al.}(2009)\citenamefont{Jia, Goswami, and
  Chakravarty}}]{Jia:2009}
\bibinfo{author}{\bibfnamefont{X.}~\bibnamefont{Jia}},
  \bibinfo{author}{\bibfnamefont{P.}~\bibnamefont{Goswami}}, \bibnamefont{and}
  \bibinfo{author}{\bibfnamefont{S.}~\bibnamefont{Chakravarty}},
  \bibinfo{journal}{Phys. Rev. B} \textbf{\bibinfo{volume}{80}},
  \bibinfo{pages}{134503} (\bibinfo{year}{2009}).

\bibitem[{\citenamefont{Dimov et~al.}(2008)\citenamefont{Dimov, Goswami, Jia,
  and Chakravarty}}]{Dimov:2008}
\bibinfo{author}{\bibfnamefont{I.}~\bibnamefont{Dimov}},
  \bibinfo{author}{\bibfnamefont{P.}~\bibnamefont{Goswami}},
  \bibinfo{author}{\bibfnamefont{X.}~\bibnamefont{Jia}}, \bibnamefont{and}
  \bibinfo{author}{\bibfnamefont{S.}~\bibnamefont{Chakravarty}},
  \bibinfo{journal}{Phys. Rev. B} \textbf{\bibinfo{volume}{78}},
  \bibinfo{pages}{134529} (\bibinfo{year}{2008}).

\bibitem[{\citenamefont{Nguyen and Chakravarty}(2002)}]{Nguyen:2002}
\bibinfo{author}{\bibfnamefont{H.~K.} \bibnamefont{Nguyen}} \bibnamefont{and}
  \bibinfo{author}{\bibfnamefont{S.}~\bibnamefont{Chakravarty}},
  \bibinfo{journal}{Phys. Rev. B} \textbf{\bibinfo{volume}{65}},
  \bibinfo{pages}{180519} (\bibinfo{year}{2002}).

\bibitem[{\citenamefont{Ghosal and Kee}(2004)}]{Ghosal:2004}
\bibinfo{author}{\bibfnamefont{A.}~\bibnamefont{Ghosal}} \bibnamefont{and}
  \bibinfo{author}{\bibfnamefont{H.-Y.} \bibnamefont{Kee}},
  \bibinfo{journal}{Physical Review B} \textbf{\bibinfo{volume}{69}},
  \bibinfo{pages}{224513} (\bibinfo{year}{2004}).

\bibitem[{\citenamefont{Kramer and Schreiber}(1996)}]{Kramer:1996}
\bibinfo{author}{\bibfnamefont{B.}~\bibnamefont{Kramer}} \bibnamefont{and}
  \bibinfo{author}{\bibfnamefont{M.}~\bibnamefont{Schreiber}}, in
  \emph{\bibinfo{booktitle}{Computational Physics}}, edited by
  \bibinfo{editor}{\bibfnamefont{K.~H.} \bibnamefont{Hoffmann}}
  \bibnamefont{and} \bibinfo{editor}{\bibfnamefont{M.}~\bibnamefont{Schreiber}}
  (\bibinfo{publisher}{Springer}, \bibinfo{address}{Berlin},
  \bibinfo{year}{1996}), p. \bibinfo{pages}{166}.

\bibitem[{\citenamefont{Pichard and Andr\'e}(1986)}]{Pichard:1986}
\bibinfo{author}{\bibfnamefont{J.~L.} \bibnamefont{Pichard}} \bibnamefont{and}
  \bibinfo{author}{\bibfnamefont{G.}~\bibnamefont{Andr\'e}},
  \bibinfo{journal}{Europhys. Lett.} \textbf{\bibinfo{volume}{2}},
  \bibinfo{pages}{477} (\bibinfo{year}{1986}).

\bibitem[{\citenamefont{Fisher and Lee}(1981)}]{Fisher:1981}
\bibinfo{author}{\bibfnamefont{D.~S.} \bibnamefont{Fisher}} \bibnamefont{and}
  \bibinfo{author}{\bibfnamefont{P.~A.} \bibnamefont{Lee}},
  \bibinfo{journal}{Phys. Rev. B} \textbf{\bibinfo{volume}{23}},
  \bibinfo{pages}{6851} (\bibinfo{year}{1981}).

\bibitem[{\citenamefont{Kumar and Jayannavar}(1992)}]{Kumar:1992}
\bibinfo{author}{\bibfnamefont{N.}~\bibnamefont{Kumar}} \bibnamefont{and}
  \bibinfo{author}{\bibfnamefont{A.~M.} \bibnamefont{Jayannavar}},
  \bibinfo{journal}{Phys. Rev. B} \textbf{\bibinfo{volume}{45}},
  \bibinfo{pages}{5001} (\bibinfo{year}{1992}).

\bibitem[{\citenamefont{Motoyama et~al.}(2007)\citenamefont{Motoyama, Yu,
  Vishik, Vajk, Mang, and Greven}}]{Motoyama:2007}
\bibinfo{author}{\bibfnamefont{E.~M.} \bibnamefont{Motoyama}},
  \bibinfo{author}{\bibfnamefont{G.}~\bibnamefont{Yu}},
  \bibinfo{author}{\bibfnamefont{I.~M.} \bibnamefont{Vishik}},
  \bibinfo{author}{\bibfnamefont{O.~P.} \bibnamefont{Vajk}},
  \bibinfo{author}{\bibfnamefont{P.~K.} \bibnamefont{Mang}}, \bibnamefont{and}
  \bibinfo{author}{\bibfnamefont{M.}~\bibnamefont{Greven}},
  \bibinfo{journal}{Nature} \textbf{\bibinfo{volume}{445}},
  \bibinfo{pages}{186} (\bibinfo{year}{2007}).

\bibitem[{\citenamefont{Nayak}(2000)}]{Nayak:2000}
\bibinfo{author}{\bibfnamefont{C.}~\bibnamefont{Nayak}},
  \bibinfo{journal}{Phys. Rev. B} \textbf{\bibinfo{volume}{62}},
  \bibinfo{pages}{4880} (\bibinfo{year}{2000}).

\bibitem[{\citenamefont{Garcia-Aldea and Chakravarty}(2010)}]{Garcia:2010}
\bibinfo{author}{\bibfnamefont{D.}~\bibnamefont{Garcia-Aldea}}
  \bibnamefont{and}
  \bibinfo{author}{\bibfnamefont{S.}~\bibnamefont{Chakravarty}},
  \bibinfo{journal}{arXiv:1008.2030}  (\bibinfo{year}{2010}).

\bibitem[{\citenamefont{Sebastian et~al.}(2009)\citenamefont{Sebastian,
  Harrison, Mielke, Liang, Bonn, Hardy, and Lonzarich}}]{Sebastian:2009}
\bibinfo{author}{\bibfnamefont{S.~E.} \bibnamefont{Sebastian}},
  \bibinfo{author}{\bibfnamefont{N.}~\bibnamefont{Harrison}},
  \bibinfo{author}{\bibfnamefont{C.~H.} \bibnamefont{Mielke}},
  \bibinfo{author}{\bibfnamefont{R.}~\bibnamefont{Liang}},
  \bibinfo{author}{\bibfnamefont{D.~A.} \bibnamefont{Bonn}},
  \bibinfo{author}{\bibfnamefont{W.~N.} \bibnamefont{Hardy}}, \bibnamefont{and}
  \bibinfo{author}{\bibfnamefont{G.~G.} \bibnamefont{Lonzarich}},
  \bibinfo{journal}{Phys. Rev. Lett.} \textbf{\bibinfo{volume}{103}},
  \bibinfo{pages}{256405} (\bibinfo{year}{2009}).

\bibitem[{\citenamefont{Sebastian et~al.}(2010)\citenamefont{Sebastian,
  Harrison, Goddard, Altarawneh, Mielke, Liang, Bonn, Hardy, Andersen, and
  Lonzarich}}]{Sebastian:2010}
\bibinfo{author}{\bibfnamefont{S.~E.} \bibnamefont{Sebastian}},
  \bibinfo{author}{\bibfnamefont{N.}~\bibnamefont{Harrison}},
  \bibinfo{author}{\bibfnamefont{P.~A.} \bibnamefont{Goddard}},
  \bibinfo{author}{\bibfnamefont{M.~M.} \bibnamefont{Altarawneh}},
  \bibinfo{author}{\bibfnamefont{C.~H.} \bibnamefont{Mielke}},
  \bibinfo{author}{\bibfnamefont{R.}~\bibnamefont{Liang}},
  \bibinfo{author}{\bibfnamefont{D.~A.} \bibnamefont{Bonn}},
  \bibinfo{author}{\bibfnamefont{W.~N.} \bibnamefont{Hardy}},
  \bibinfo{author}{\bibfnamefont{O.~K.} \bibnamefont{Andersen}},
  \bibnamefont{and} \bibinfo{author}{\bibfnamefont{G.~G.}
  \bibnamefont{Lonzarich}}, \bibinfo{journal}{Phys. Rev. B}
  \textbf{\bibinfo{volume}{81}}, \bibinfo{pages}{214524}
  (\bibinfo{year}{2010}).

\bibitem[{\citenamefont{Ramshaw et~al.}(2010)\citenamefont{Ramshaw, Vignolle,
  Liang, Hardy, Proust, and Bonn}}]{Ramshaw:2010}
\bibinfo{author}{\bibfnamefont{B.~J.} \bibnamefont{Ramshaw}},
  \bibinfo{author}{\bibfnamefont{B.}~\bibnamefont{Vignolle}},
  \bibinfo{author}{\bibfnamefont{R.}~\bibnamefont{Liang}},
  \bibinfo{author}{\bibfnamefont{W.~N.} \bibnamefont{Hardy}},
  \bibinfo{author}{\bibfnamefont{C.}~\bibnamefont{Proust}}, \bibnamefont{and}
  \bibinfo{author}{\bibfnamefont{D.~A.} \bibnamefont{Bonn}},
  \bibinfo{journal}{arXiv:1004.0260}  (\bibinfo{year}{2010}).

\end{thebibliography}
\end{document}